\documentclass[useAMS,usenatbib,usegraphicx,fleqn]{mn2e}
\usepackage{amsmath}

\usepackage[normalem]{ulem}

\begin{document}
\title[Isotropy of the Cosmic Expansion]{The Hubble Expansion is Isotropic in the Epoch of Dark Energy}

\author[Darling]{ Jeremy Darling$^1$\\
$^1$Center for Astrophysics and Space Astronomy,
Department of Astrophysical and Planetary Sciences,
University of Colorado, \\389 UCB, Boulder, CO 80309-0389, USA; jdarling@colorado.edu}

\maketitle
\begin{abstract}
The isotropy of the universal Hubble expansion is a fundamental tenet of physical cosmology, 
but it has not been precisely tested during the current epoch, when dark energy is dominant.  
Anisotropic expansion will produce a shearing velocity field, causing objects to 
stream toward directions of faster expansion and away from directions of slower expansion.
This work tests the basic cosmological assumption of isotropic expansion and thus the isotropy 
of  dark energy.
The simplest anisotropy will manifest as a quadrupolar curl-free proper motion vector field.  
We derive this theoretical signature using a tri-axial expanding metric 
with a flat geometry (Bianchi \rm{I} model), generalizing and correcting previous work.  We then employ 
the best current data, the Titov \& Lambert (2013) proper motion catalog of 429 objects, to measure the isotropy of 
universal expansion.  
We demonstrate that the Hubble expansion is isotropic to 7\% ($1\,\sigma$), corresponding to streaming motions of 
1 microarcsecond yr$^{-1}$, in the best-constrained directions
($-$19\% and +17\% in the least-constrained directions) and does
not significantly deviate from isotropy in any direction. 
The Gaia mission, which is expected to obtain proper motions for 500,000 quasars, 
will likely constrain the anisotropy below 1\%.
\end{abstract}

\section{Introduction}


The isotropy of the cosmic expansion is well-constrained for the early universe, particularly by Cosmic Microwave
Background observations, and is a basic tenet of physical cosmology.  The change from a matter-dominated to a 
dark energy-dominated universe in recent times, however, raises the possibility of a dark energy-driven
anisotropic expansion if dark energy is itself anisotropic.  There is no obvious reason for such symmetry 
breaking, but observational tests of something as fundamental as the isotropy of the Hubble expansion 
should be made for late times (the current epoch).  

One such test is possible via extragalactic proper motions:  if the expansion is anisotropic, then quasars and 
galaxies will stream toward directions of faster expansion and away from directions of slower
expansion.  The signature of anisotropic expansion in a homogeneous universe is thus a curl-free proper motion 
vector field \citep[to first order;][]{quercellini09,fontanini09,titov09}.  

The term ``cosmic parallax'' has been used by some to indicate a general relative angular motion
of objects in the universe \citep[e.g.,][]{quercellini09,fontanini09} 
and used by others in a more canonical sense to indicate apparent angular motion induced by the motion 
of the observer \citep[e.g.,][]{ding09}.  We favor the latter usage; this paper therefore treats the apparent proper 
motion induced by anisotropic cosmic expansion \citep{amendola13}, referenced to the International 
Celestial Reference Frame (ICRF) in the current epoch.  The observed proper motions are therefore relative, 
but are not necessarily induced by the observer's motion.

In this paper, we present a simple model of anisotropic expansion and 
fit the model to the \citet{titov13} proper motion catalog to place a new
constraint on the isotropy of the Hubble expansion and thus on the isotropy of dark energy.   
We assume $H_\circ=72$~km~s$^{-1}$~Mpc$^{-1}$
and a flat cosmology (this treatment is independent of specific assumptions about
$\Omega_\Lambda$ and $\Omega_M$, provided $\Omega_\Lambda + \Omega_M = 1$).

\section{Anisotropic Expansion Model}\label{sec:theory}

As described in \citet{quercellini09} and \citet{fontanini09}, a homogeneous but anisotropic Bianchi \rm{I} model with metric 
\begin{equation}
  ds^2 = -dt^2+a^2(t)\,dx'^{\,2}+b^2(t)\,dy'^{\,2}+c^2(t)\,dz'^{\,2}  \label{eqn:metric}
\end{equation}
has three different expansion rates, $H_{x'} = \dot{a}/a$, $H_{y'} = \dot{b}/b$, and $H_{z'} = \dot{c}/c$, 
where the Hubble parameter as observed is $H={d\over dt}(abc)^{1/3}/(abc)^{1/3}$ and the Friedmann-Robertson-Walker
metric is recovered for $a(t)=b(t)=c(t)$.  This metric has a flat geometry and no global vorticity, but the 
anisotropic expansion will produce a shearing velocity field, causing objects to
stream toward directions of faster expansion and away from directions of slower expansion.  The shear 
can be characterized by the fractional deviation from the average Hubble expansion today ($t=t_\circ$), 
\begin{equation}
  \Sigma_{i'} = {H_{i',\circ}\over H_\circ}-1,   \label{eqn:sheardefn}
\end{equation}
where $i' = x'$, $y'$, or $z'$, isotropy corresponds to $\Sigma_{x'} = \Sigma_{y'} = \Sigma_{z'} =0$
(no deviation from $H_\circ$ in any direction), and the expansion is ``conserved'':
$\Sigma_{x'}+\Sigma_{y'}+\Sigma_{z'}=0$ (the directionless overall $H_\circ$ is preserved, despite anisotropy).
Under the simplifying assumption of straight geodesics
(incorrect, but a small error as demonstrated by \citealt{quercellini09}),
the sky signal of an anisotropic expansion is a curl-free quadrupolar proper motion vector field that is 
independent of distance. Note that this is a ``real-time'' signal, meaning that the time derivatives are with 
respect to small coordinate time intervals today (decades) and therefore a constant $H_\circ$ is a good
assumption.

The application of this model by \citet{quercellini09},
\citet{titov09}, and \citet{titov11} to the apparent proper motions of extragalactic objects, 
however, retained a metric fixed with respect to the equatorial coordinate
system and did not allow the anisotropy to have an arbitrary orientation.  Application of this model to 
observations by \citet{titov11} was not rotationally invariant (the result would depend on the choice of coordinate
system).  In order to obtain a fully general model, we require rotational invariance appropriate for 
vector spherical harmonics (the divergence and curl of the scalar spherical harmonics, $Y_{\ell m}$):  
all orders $m$ for a given degree $\ell$ must be included in a model fit to data \citep{mignard12}.  
Here we demonstrate that all quadrupolar ($\ell=2$) curl-free vector spherical harmonic orders naturally arise from an
arbitrary rotation of the anisotropic coordinate system.  

Using an arbitrary rotation from the equatorial reference frame to the one described by Equation (\ref{eqn:metric}), 
we can allow for any anisotropy orientation.  The anisotropy also need not be triaxial; there could simply be a 
single direction of high or low expansion, provided the expansion conservation condition is satisfied.  
Rotations are made about $\hat{z}$ by angle $\alpha^*$, about the new $\hat{y}$ by $\delta^*$, and about the final 
$\hat{x}'$ axis by $\psi^*$.  Thus, the anisotropic Bianchi \rm{I} axes can be expressed in the equatorial coordinate system
via $\bmath{x}' = \mathbfss{R}_x(\psi^*)\,\mathbfss{R}_y(\delta^*)\, \mathbfss{R}_z(\alpha^*)\, \bmath{x}$, where $\mathbfss{R}_i(\phi)$ is 
the rotation matrix about axis $i$ by angle $\phi$.  
The equatorial coordinates of the anisotropy axes are therefore:
\begin{subequations}
\begin{eqnarray}
(\alpha_{x'},\delta_{x'}) = (\alpha^*,\delta^*) \label{eqn:xaxis} 
\end{eqnarray}
\begin{eqnarray} 
 (\alpha_{y'},\delta_{y'}) = \left({\sin^{-1}(\cos\alpha^*\cos\psi^*-\sin\alpha^*\sin\delta^*\sin\psi^*)\over\sqrt{1-\cos^2\delta^*\sin^2\psi^*}}\, ,
 \right. \nonumber \\ 
  \left. \sin^{-1}(\cos\delta^*\sin\psi^*) \vphantom{\sin^{-1}(\cos\alpha^*)\over\sqrt{\cos^2\delta^*}} \right)  \label{eqn:yaxis}
\end{eqnarray}
\begin{eqnarray} 
(\alpha_{z'},\delta_{z'}) = \left({\sin^{-1}(-\cos\alpha^*\sin\psi^*-\sin\alpha^*\sin\delta^*\cos\psi^*)\over\sqrt{1-\cos^2\delta^*\cos^2\psi^*}}\, ,
  \right. \nonumber \\ \left.
   \sin^{-1}(\cos\delta^*\cos\psi^*) \vphantom{\sin^{-1}(\cos\alpha^*)\over\sqrt{\cos^2\delta^*}} \right).  \label{eqn:zaxis} 
\end{eqnarray}
\end{subequations}
 
The anisotropic expansion proper motion vector field $\bmath{V}_{\rm Shear}(\alpha,\delta)$ coefficients are listed in 
Table \ref{tab:shear_theory} (see Appendix \ref{appdx} for the full equation).  
These were obtained by taking the time derivatives of the equatorial coordinates, $\dot{\alpha}\cos\delta$ and $\dot{\delta}$,
expressed in terms of the shear terms $\Sigma_{i'}$.  
The general form of the vector field is 
\begin{eqnarray}
 \bmath{V}_{\rm Shear}(\alpha,\delta) = H_\circ\ \sum_{m=0}^\ell\ \sum_{i=\alpha,\delta}\ \xi_{2m,i}^{Re,Im}(\alpha,\delta) \nonumber \\
           \times\left[ a_{2m}^{Re,Im}(\alpha^*,\delta^*,\psi^*) \left(\Sigma_{y'}+{1\over2}\,\Sigma_{x'}\right) \right. \nonumber \\
            \left.    +\  b_{2m}^{Re,Im}(\alpha^*,\delta^*,\psi^*) \left( {1\over2}\,\Sigma_{x'}\right)\right] \bmath{\hat{e}}_i
\label{eqn:sheareqn}
\end{eqnarray} 
where there are no imaginary coefficients for $m=0$ and the sum over real and imaginary coefficients is implied.  
The Hubble constant can be written as $H_\circ = 15.2$~$\mu$as~yr$^{-1}$
(for $H_\circ = 72$~km~s$^{-1}$~Mpc$^{-1}$), so a shear of 10\% would produce streaming motions of order 1.5~$\mu$as~yr$^{-1}$.

Comparison of this derivation of the shear field $\bmath{V}_{\rm Shear}(\alpha,\delta)$ to the spheroidal 
(curl-free or E-mode) quadrupolar vector spherical harmonics \citep{mignard12}
reveals an exact one-to-one correspondence (Table \ref{tab:shear_theory}).  
The treatment here is therefore rotationally invariant (as desired), and it is therefore completely equivalent to 
fit the spheroidal $\ell=2$ vector spherical harmonics to a proper motion vector field to test the isotropy of expansion.   
It is not, however, correct to fit or select single or selected orders of a vector spherical harmonic model, as has been done 
previously.  Table \ref{tab:shear_theory} lists the equivalent vector spherical harmonic coefficients (see Appendix \ref{appdx}
for the full equation), following the \citet{mignard12} conventions:
\begin{eqnarray}
 \bmath{V}_{E2}(\alpha,\delta) = \sum_{m=0}^\ell\ \sum_{i=\alpha,\delta}\ \xi_{2m,i}^{Re,Im}(\alpha,\delta)\,\chi_{2m}^{Re,Im} 
       s_{2m}^{Re,Im}\,\bmath{\hat{e}}_i\ .
\label{eqn:E2eqn}
\end{eqnarray} 
Note that this equation has absorbed the factors of 2 for the $m>0$ orders and the factors of $-1$ for the imaginary terms
into the coefficients $\chi_{2m}^{Re,Im}$, in contrast to the definitions used by \citet{mignard12}.  

\begin{table*}
\begin{minipage}{135mm}
\caption{Anisotropy Model Coefficients}
\label{tab:shear_theory}
\begin{tabular}{@{}cccccc@{}}
\hline
$\bmath{\hat{e}}_\alpha$ &
$\bmath{\hat{e}}_\delta$  &
$\Sigma_{y'}+{1\over2}\Sigma_{x'}$ &
${1\over2}\Sigma_{x'}$ &
\multicolumn{2}{c}{$\bmath{V}_{E2}$} \\  \noalign{\vskip 2mm} 
$\xi_{2m,\alpha}^{Re,Im}(\alpha,\delta)$ & 
$\xi_{2m,\delta}^{Re,Im}(\alpha,\delta)$ & 
$a_{2m}^{Re,Im}(\alpha^*,\delta^*,\psi^*)$ & 
$b_{2m}^{Re,Im}(\alpha^*,\delta^*,\psi^*)$ &
$s_{2m}$ & 
$\chi_{2m}^{Re,Im}$ 
\\
\hline
0 & ${3\over8}\sin2\delta$ & $-\cos2\psi^*(1+\cos2\delta^*)$ & $1-3\cos2\delta^*$ & $s_{20}$ & ${2\over3}\sqrt{15\over2\pi}$ \\ 
${1\over2} \sin\alpha\sin\delta$ & $-{1\over2}\cos\alpha\cos2\delta$ & \vtop{\hbox{\strut $2\sin\alpha^*\cos\delta^*\sin2\psi^*$}\hbox{\strut $-\cos\alpha^*\sin2\delta^*\cos2\psi^*$}} & $-3\cos\alpha^*\sin2\delta^*$ & $s_{21}^{Re}$ &  $\sqrt{5\over\pi}$\\ 
${1\over2} \cos\alpha\sin\delta$ & ${1\over2}\sin\alpha\cos2\delta$ & \vtop{\hbox{\strut $2\cos\alpha^*\cos\delta^*\sin2\psi^*$}\hbox{\strut $+\sin\alpha^*\sin2\delta^*\cos2\psi^*$}} & $3\sin\alpha^*\sin2\delta^*$ & $s_{21}^{Im}$ & $\sqrt{5\over\pi}$\\ 
${1\over4} \sin2\alpha\cos\delta$ & ${1\over8}\cos2\alpha\sin2\delta$ & \vtop{\hbox{\strut $3\cos2\alpha^*\cos2\psi^*$}\hbox{\strut $-\cos2\alpha^*\cos2\delta^*\cos2\psi^*$}\hbox{\strut $-4\sin2\alpha^*\sin\delta^*\sin2\psi^*$}} & $-3\cos2\alpha^*(1+\cos2\delta^*)$ & $s_{22}^{Re}$ & $-2\sqrt{5\over\pi}$ \\ 
${1\over4} \cos2\alpha\cos\delta$ & $-{1\over8}\sin2\alpha\sin2\delta$ & \vtop{\hbox{\strut $-3\sin2\alpha^*\cos2\psi^*$}\hbox{\strut $+\sin2\alpha^*\cos2\delta^*\cos2\psi^*$}\hbox{\strut $-4\cos2\alpha^*\sin\delta\sin2\psi^*$}} & $3\sin2\alpha^*(1+\cos2\delta^*)$ & $s_{22}^{Im}$ & $-2\sqrt{5\over\pi}$ \\
\hline
\end{tabular}

Model coefficients and angular terms corresponding to the terms in Equations (\ref{eqn:sheareqn}) and (\ref{eqn:E2eqn}).
See Equation (\ref{eqn:fullshear}) for the full shear vector field and Equation (\ref{eqn:E2})
for the full spheroidal quadrupolar vector field.
\end{minipage}
\end{table*}

\section{Data Analysis Methods}\label{sec:methods}

We employ the \citet{titov13} proper motion measurements of 429 radio sources to examine the isotropy of the Hubble 
expansion.   The data were obtained from sessions of the permanent geodetic and astrometric VLBI program, which includes 
the Very Long Baseline Array\footnote{The National Radio Astronomy Observatory is a facility of the National Science 
Foundation operated under cooperative agreement by Associated Universities, Inc.}, at 8.4 GHz in 1990--2013 using a 
relaxed per-session no-net rotation constraint and an iterative process to reject objects with large intrinsic or spurious proper motions
\citep{titov13}.
Objects in this catalog can show large intrinsic proper motions due to plasmon ejection in jets or due to core
shift effects, but these motions are uncorrelated between objects; they simply add intrinsic proper motion noise
to any correlated global signals.  

\citet{titov11} first detected the secular aberration drift quasar proper motion signature induced by the
barycenter acceleration about the Galactic Center, which was later confirmed by \citet{xu12} and 
refined by \citet{titov13}.  The signature is an E-mode (curl-free) proper motion dipole 
with apex at the Galactic Center.  In order to measure or constrain the E-mode quadrupolar anisotropy 
signal, we subtract the dipole proper motion pattern from the observed proper motion vector field, 
but employ the \citet{reid14} results obtained from 
trigonometric parallaxes and proper motions of masers associated with young massive stars:
they obtain a Galactic Center distance of $R_0=8.34\pm0.16$~kpc and 
a rotation speed at $R_0$ of $\Theta_0=240\pm8$~km~s$^{-1}$.  Since the relevant quantity 
for aberration drift is the solar
acceleration about the Galactic Center, we use the 
actual solar orbital motion that includes the solar motion in the direction of the 
Galactic rotation,  $\Theta_0+V_\odot=255.2\pm5.1$~km~s$^{-1}$ \citep{reid14},
which yields an acceleration of $0.80\pm0.04$~cm~s$^{-1}$~yr$^{-1}$ and
a dipole amplitude of $5.5\pm0.2$~$\mu$as~yr$^{-1}$.
This dipole in the \citet{titov13} notation is 
$\vec{d} = (d_1,d_2,d_3) =  (-0.300\pm0.013,-4.80\pm0.21,-2.66\pm0.12)$~$\mu$as~yr$^{-1}$; 
in the \cite{mignard12} 
notation, it is $(s_{10},s_{11}^{Re},s_{11}^{Im}) = (-7.71\pm0.34,+0.615\pm0.027,-9.82\pm0.44)$~$\mu$as~yr$^{-1}$.
We assume that the acceleration direction is exactly toward the Galactic Center and 
do not include the out-of-the-disk acceleration described by \citet{xu12} in our correction
(following \citealt{darling13}).
\citet{titov13} do not confirm the acceleration detected by \citet{xu12}.  

The derived dipole has substantially smaller errors than the \citet{titov13} measurement, and when we
subtract this dipole from the vector field, we introduce negligible statistical errors compared to the 
proper motion uncertainty in individual objects.  Although dipole and quadrupole signals are in principle
orthogonal, covariance between different-degree vector spherical harmonics do exist \citep[e.g.,][]{titovmalkin09,titov13}, so
subtraction of the best-measured dipole --- using completely independent observations --- is appropriate before measuring 
the E-mode quadrupole.

After subtracting the dipole signal from the quasar proper motion catalog, we perform a least-squares minimization
fit of the observed proper motions to the anisotropy model described by Equations (\ref{eqn:sheareqn}) and 
(\ref{eqn:fullshear}) and Table \ref{tab:shear_theory}.  The free parameters are the rotation angles between 
the equatorial reference frame and the anisotropy frame, $\alpha^*$, $\delta^*$, and $\psi^*$, and two of the 
shear parameters describing the anisotropy, $\Sigma_{x'}$ and $\Sigma_{y'}$.  The third shear parameter, 
$\Sigma_{z'}$, is determined by the expansion conservation condition (Section \ref{sec:theory}).  The Hubble constant 
is assumed.  Unless genuine significant anisotropy is detected, this model does not constrain $H_\circ$, which
acts as a scaling amplitude for the streaming proper motions ($H_\circ = 15.2$~$\mu$as~yr$^{-1}$).  
 We also fit the spheroidal quadrupole vector spherical harmonics for comparison.

\section{Results}\label{sec:results}

Table \ref{tab:shear_meas} shows the measured anisotropy based on the least-squares fitting of the
proper motion catalog to the anisotropy model.  
Table \ref{tab:E2quad_meas} shows the 
more general vector spherical harmonic parameters for a spheroidal quadrupole proper motion vector field.
None of the shear parameters nor the quadrupole vector spherical harmonic coefficients are significant; all 
are consistent with zero, indicating an isotropic Hubble expansion.  The largest deviation from isotropy is
$-19\%\pm7\%$, and the largest positive deviation is $+17\%\pm7\%$ (neither significant).  The smallest deviation 
(and thus the best anisotropy constraint) is $+2\%\pm7\%$.  The anisotropy of the Hubble expansion in the epoch of 
dark energy is thus less than 7\% ($1\,\sigma$) in the best-constrained direction. 
\begin{table}
\caption{Measured Expansion Anisotropy}
\label{tab:shear_meas}
\begin{tabular}{@{}cccccc@{}}
\hline
$\Sigma_{x'}$ &
$\Sigma_{y'}$ &
$\Sigma_{z'}$ &
$\alpha^*$($^\circ$) &
$\delta^*$($^\circ$) &
$\psi^*$($^\circ$) \\
\hline
0.17(7) & $-$0.19(7) & 0.02(7) & 193(15) & 47(26) & $-$2(21)\\
\hline
\end{tabular}

\smallskip
The expansion shear terms $\Sigma_{i'}$ indicate the fractional 
departure from the average Hubble expansion (Equation (\ref{eqn:sheardefn})). The 
$\hat{x}'$ axis lies in the $(\alpha^*,\delta^*)$ direction, and the $\hat{y}'$ and $\hat{z}'$ 
axes are rotated about the $\hat{x}'$ axis by $\psi^*$; their coordinates are listed generally 
in Equations (\ref{eqn:yaxis}) and (\ref{eqn:zaxis}) and as measured in Section \ref{sec:results}. 
Parenthetical values are $1\,\sigma$ uncertainties on the final digit(s).  
\end{table}
\begin{table}
\caption{Measured Spheroidal Quadrupole}
\label{tab:E2quad_meas}
\begin{tabular}{@{}cccccc@{}}
\hline
$s_{20}$ & $s_{21}^{Re}$ & $s_{21}^{Im}$ & $s_{22}^{Re}$ & $s_{22}^{Im}$ & $\sqrt{P_2^s}$\\ 
\hline 
3.0(2.5)  & 1.7(1.4) &  $-$0.5(1.7) & 3.1(1.4) & $-$1.4(1.3) & 6.2(2.1)\\
\hline 
\end{tabular}

\smallskip
The quadrupolar ($\ell=2$) spheroidal vector spherical harmonic coefficients $s_{2m}$ and 
the total power in the $\ell=2$ curl-free order $P_2^s$ follow the \citet{mignard12} conventions.  
Parenthetical values are $1\,\sigma$ uncertainties. The units are $\mu$as~yr$^{-1}$.
The Z-score of a one-sided significance test
is 1.1, so the power in this mode is not significant.  
\end{table}

\begin{figure}
\includegraphics[width=0.48\textwidth]{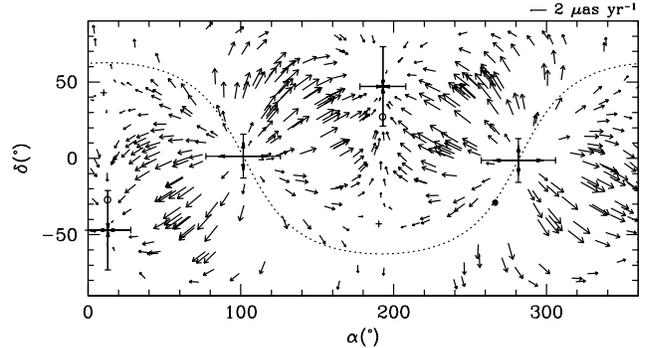}
\caption{
The best-fit model anisotropy vector field in equatorial coordinates.
The model parameters and uncertainties are listed in Tables \ref{tab:shear_meas} and  \ref{tab:E2quad_meas}.  
The measured anisotropy is not significant, either by parameter in or total power.  
The high and low deviations from the average Hubble expansion are indicated as converging ($\Sigma_{x'}=0.17\pm0.07$)
and diverging ($\Sigma_{y'}=-0.19\pm0.07$) loci, including error bars.
The upper right scale bar indicates the amplitude of the vectors, the dotted line shows the Galactic plane, the solid circle
indicates the Galactic Center, the open circles mark the Galactic poles, and the crosses indicate the $\hat{z}'$ direction.}
\label{fig:shear}
\end{figure}

Figure \ref{fig:shear} shows the model fit to the proper motion vector field for the 429 objects in the \citet{titov13}
catalog.  Positive deviation from the Hubble expansion ($\Sigma_{i'} > 0$) appears as an antipodal pair of convergent points, 
and negative deviation appears as an antipodal pair of divergent points.  
The equatorial coordinates of the best-fit (but not significant) anisotropy axes are:
$(\alpha_{x'},\delta_{x'}) = (13^\circ\pm15^\circ,-47^\circ\pm26^\circ)$  and $(193^\circ\pm15^\circ,+47^\circ\pm26^\circ)$,
$(\alpha_{y'},\delta_{y'}) = (102^\circ\pm24^\circ,+1^\circ\pm14^\circ)$ and $(282^\circ\pm24^\circ,-1^\circ\pm14^\circ)$, and
$(\alpha_{z'},\delta_{z'}) = (11^\circ\pm33^\circ,+43^\circ\pm26^\circ)$ and $(191^\circ\pm33^\circ,-43^\circ\pm26^\circ)$.  
These results are insensitive to the initial parameter assumptions.  While the anisotropy axes can be 
interchanged arbitrarily through rotations, the shear values and directions are stable 
best-fit solutions.  

The solution is likewise insensitive to the sometimes large intrinsic proper motions or large proper motion 
uncertainties of individual objects.  Restricting the sample to the 358 objects with 
proper motions and proper motion uncertainties less than 100~$\mu$as~yr$^{-1}$ has a negligible impact
on the model parameters or uncertainties listed in Table \ref{tab:shear_meas}.

The power in the spheroidal vector spherical harmonics of degree $\ell$ is 
\begin{equation}
 P_\ell^s = s_{\ell0}^2+2 \sum_{m=1}^\ell \left(\left(s_{\ell m}^{Re}\right)^2+\left(s_{\ell m}^{Im}\right)^2\right),
\label{eqn:power}
\end{equation}
a scalar under coordinate rotation \citep{mignard12}.  For the $\ell=2$ fit in Table \ref{tab:shear_meas}, the power
is $\sqrt{P_2^s} = 6.2\pm2.1$~$\mu$as~yr$^{-1}$.  The Z-score of a one-sided significance test 
of the mean-error-reduced power \citep[][Equations (85) and (87)]{mignard12}, is 1.1, so the power in this
mode is not significant.   

\section{Discussion}

While the Bianchi \rm{I} metric model assumed for the data analysis is specific, the fitting function is 
a general $\ell=2$ spheroidal vector spherical harmonic; the model plays a role in specific parameter estimation,
but the vector field sky pattern is generic.  One can therefore adapt the fit parameters to different
cosmological models, and one can fit additional vector spherical harmonic terms \citep[e.g.,][]{mignard12,titov13}.  

Figure \ref{fig:shear} shows a remarkable (but non-significant) alignment of the fit anisotropy axes 
with the Galactic plane-equator intersection.  This 
could be driven by a vertical spheroidal dipole component induced by vertical (out-of-the-plane) 
solar acceleration \citep[a second aberration drift;][]{xu12}.  A dipole fit to the proper motion vector field (after subtracting
the galactocentric acceleration described in Section \ref{sec:methods}) is not significant and does
not point toward the $\hat{x}'$ axis (the points of convergence).  Since the anisotropy is neither significant in total
power nor in individual parameters, this surprising alignment seems to be coincidental and should not be 
over-interpreted.  

The main limitation to this technique is the proper motion precision and sample size.  The 
overall proper motion precision of individual objects will improve with time, as will the 
sample size of ICRF ``defining sources,'' but these will be slow, secular improvements.  
The next order-of-magnitude improvement will be provided by the Gaia mission.
Gaia is an optical astrometry mission that will measure 500,000 quasar proper motions
with $\sim$80~$\mu$as astrometry  for $V=18$ mag stars \citep{debruijne05}.  Unlike radio
sources, compact optical extragalactic sources do not show significant internal intrinsic proper motions, 
so Gaia proper motion catalogs will not exhibit the uncorrelated proper motion signals that 
contaminate radio measurements.  We estimate that the Gaia mission will constrain anisotropy 
below 1\%.

\section{Conclusions}

We have demonstrated how anisotropic Hubble expansion can be measured or constrained using
extragalactic proper motions, and we applied this technique to the best current proper motion 
catalog \citep{titov13} to place a new
constraint on the isotropy of the Hubble expansion and thus on the isotropy of dark energy.   
No significant anisotropy was detected; the 
Hubble expansion is isotropic to 7\% ($1\,\sigma$), corresponding to streaming motions of 
1~$\mu$as~yr$^{-1}$, in the best-constrained directions
($-$19\% and +17\% in the least-constrained directions) and does
not significantly deviate from isotropy in any direction. 
The Gaia mission, which is expected to obtain proper motions for 500,000 quasars, 
will likely constrain the anisotropy below 1\%.

\section*{Acknowledgments}
The author thanks the anonymous referee for helpful comments and 
O. Titov and S. B. Lambert for making their proper motion catalog publicly available. 
This research has made use of the VizieR catalogue access tool, CDS, Strasbourg, France. 
The original description of the VizieR service was published in A\&AS 143, 23.
This research has made use of NASA's Astrophysics Data System Bibliographic Services.

\clearpage

\appendix

\section{Anisotropic Expansion Vector Fields}\label{appdx}

The complete vector field described in Section \ref{sec:theory}, Equation (\ref{eqn:sheareqn}), and Table \ref{tab:shear_theory}  is
\begin{eqnarray}
  \bmath{V}_{\rm{Shear}} (\alpha,\delta) = H_\circ
      \left( {1\over2} \sin\alpha\sin\delta \left[  \left(\Sigma_{y'}+{1\over2}\,\Sigma_{x'}\right) \left(2\sin\alpha^*\cos\delta^*\sin2\psi^*
     -\cos\alpha^*\sin2\delta^*\cos2\psi^*\right)
      -{3\over2}\,\Sigma_{x'}\,\cos\alpha^*\sin2\delta^*\right] \right. \nonumber\\
     \left. +{1\over2} \cos\alpha\sin\delta \left[  \left(\Sigma_{y'}+{1\over2}\,\Sigma_{x'}\right) \left(2\cos\alpha^*\cos\delta^*\sin2\psi^*+\sin\alpha^*\sin2\delta^*\cos2\psi^*\right)
                +{3\over2}\,\Sigma_{x'}\,\sin\alpha^*\sin2\delta^*\right] \right. \nonumber\\
     \left. +{1\over4} \sin2\alpha\cos\delta \left[  \left(\Sigma_{y'}+{1\over2}\,\Sigma_{x'}\right) \left(3\cos2\alpha^*\cos2\psi^*-\cos2\alpha^*\cos2\delta^*\cos2\psi^*     -4\sin2\alpha^*\sin\delta^*\sin2\psi^*\right) 
\right.\right. \nonumber\\  \left.\left. 
                                                                    -{3\over2}\,\Sigma_{x'}\,\cos2\alpha^*\left(1+\cos2\delta^*\right)\right] \right.\nonumber\\
      \left. +{1\over4} \cos2\alpha\cos\delta \left[  \left(\Sigma_{y'}+{1\over2}\,\Sigma_{x'}\right) \left(-3\sin2\alpha^*\cos2\psi^*+\sin2\alpha^*\cos2\delta^*\cos2\psi^* -4\cos2\alpha^*\sin\delta^*\sin2\psi^*\right)
\right.\right. \nonumber\\    \left. \left.                                                                                           
                                                                            +{3\over2}\,\Sigma_{x'}\,\sin2\alpha^*\left(1+\cos2\delta^*\right)\right]\right)
           \bmath{\hat{e}}_\alpha \nonumber\\
      + H_\circ \left({3\over8}\sin2\delta\left[-\left(\Sigma_{y'}+{1\over2}\,\Sigma_{x'}\right)\cos2\psi^*\left(1+\cos2\delta^*\right)
                                                                    +{1\over2}\,\Sigma_{x'}\left(1-3\cos2\delta^*\right)\right] \right. \nonumber\\
   \left.+{1\over2}\cos\alpha\cos2\delta\left[-\left(\Sigma_{y'}+{1\over2}\,\Sigma_{x'}\right)\left(2\sin\alpha^*\cos\delta^*\sin2\psi^*-\cos\alpha^*\sin2\delta^*\cos2\psi^*\right)
                                                       +{3\over2}\,\Sigma_{x'}\,\cos\alpha^*\sin2\delta^*\right] \right. \nonumber\\
     \left.   +{1\over2}\sin\alpha\cos2\delta\left[\left(\Sigma_{y'}+{1\over2}\,\Sigma_{x'}\right)\left(2\cos\alpha^*\cos\delta^*\sin2\psi^*+\sin\alpha^*\sin2\delta^*\cos2\psi^*\right)
                                                            +{3\over2}\,\Sigma_{x'}\,\sin\alpha^*\sin2\delta^*\right] \right. \nonumber\\
    \left.     +{1\over8}\cos2\alpha\sin2\delta\left[\left(\Sigma_{y'}+{1\over2}\,\Sigma_{x'}\right)\left(3\cos2\alpha^*\cos2\psi^*-\cos2\alpha^*\cos2\delta^*\cos2\psi^*                                                       -4\sin2\alpha^*\sin\delta^*\sin2\psi^*\right)
\right.\right. \nonumber\\    \left.\left.
                                               -{3\over2}\,\Sigma_{x'}\,\cos2\alpha^*\left(1+\cos2\delta^*\right)\right] \right. \nonumber\\
    \left.      +{1\over8}\sin2\alpha\sin2\delta\left[-\left(\Sigma_{y'}+{1\over2}\,\Sigma_{x'}\right)\left(-3\sin2\alpha^*\cos2\psi^*+\sin2\alpha^*\cos2\delta^*\cos2\psi^*       -4\cos2\alpha^*\sin\delta\sin2\psi^*\right)
\right.\right.\nonumber\\   \left.\left.                                                                 
                                                              -{3\over2}\,\Sigma_{x'}\, \sin2\alpha^*\left(1+\cos2\delta^*\right) \right] \right)
       \bmath{\hat{e}}_\delta.\ \ 
\label{eqn:fullshear}
\end{eqnarray} 
The E-mode (curl-free) quadrupole vector spherical harmonic is
\begin{eqnarray}
  \bmath{V}_{E2} (\alpha,\delta) = 
      \left(s_{21}^{Re}\, {1\over2} \sqrt{5\over\pi}\, \sin\alpha\sin\delta+s_{21}^{Im}\, {1\over2} \sqrt{5\over\pi}\, \cos\alpha\sin\delta
- s_{22}^{Re}\, {1\over2} \sqrt{5\over\pi}\, \sin2\alpha\cos\delta 
-s_{22}^{Im}\, {1\over2} \sqrt{5\over\pi}\, \cos2\alpha\cos\delta\right) \bmath{\hat{e}}_\alpha \nonumber\\
     + \left(s_{20}\, {1\over4}\sqrt{15\over 2\pi}\, \sin2\delta - s_{21}^{Re}\, {1\over2} \sqrt{5\over\pi}\, \cos\alpha\cos2\delta
     + s_{21}^{Im}\, {1\over2} \sqrt{5\over\pi}\, \sin\alpha\cos2\delta 
     - s_{22}^{Re}\, {1\over4}\sqrt{5\over\pi}\, \cos2\alpha\sin2\delta 
\right.\nonumber\\ \left. 
      + s_{22}^{Im}\, {1\over4}\sqrt{5\over\pi}\,
      \sin2\alpha\sin2\delta \right) \bmath{\hat{e}}_\delta.\ \ 
\label{eqn:E2}
\end{eqnarray}


\begin{thebibliography}{}
\bibitem[Amendola et al.(2013)]{amendola13}  Amendola, L., Eggers B. O., Valkenburg, W., \& Wong, Y. Y. Y.  2013, 
  Journal of Cosmology and Astroparticle Physics, 12, 042
\bibitem[Darling(2013)]{darling13}  Darling, J.  2013, ApJ, 777, L21
\bibitem[de Bruijne et al.(2005)]{debruijne05}  de Bruijne, J., Perryman, M., Lindegren, L., Jordi, C., H\o g, E., Katz, D., \& Cropper, M.
2005, Gaia-JdB-022 Technical Note
\bibitem[Ding \& Croft(2009)]{ding09}  Ding, F. \& Croft, R. A. C.   2009, MNRAS, 397, 1739
\bibitem[Fontanini et al.(2009)]{fontanini09} Fontanini, M., West, E. J., \& Trodden, M.  2009, Phys.\ Rev.\ D, 80, 123515
\bibitem[Mignard \& Klioner(2012)]{mignard12}  Mignard, F. \& Klioner, S.  2012, A\&A, 547, A59
\bibitem[Quercellini et al.(2009)]{quercellini09} Quercellini, C., Cabella, P., Amendola, L., Quartin, M., \& Balbi, A.  
2009, Phys. Rev. D, 80, 063527
\bibitem[Reid et al.(2014)]{reid14}  Reid, M. J., Menten, K. M., Brunthaler, A., Zheng, X. W., Dame, T. M., Xu, Y., Wu, Y., Zhang, B., Sanna, A., Sato, M., Hachisuka, K., Choi, Y. K., Immer, K., Moscadelli, L., Rygl, K. L. J., \& Bartkiewicz, A.  2014, in press (arXiv 1401.5377)
\bibitem[Titov(2009)]{titov09} Titov, O.  2009, in Proc. 19th European VLBI for Geodesy and Astrometry (EVGA) Working Meeting, ed. G. Bourda, et al., 14
\bibitem[Titov \& Malkin(2009)]{titovmalkin09}  Titov, O. \& Malkin, Z.  2009, A\&A, 506, 1477
\bibitem[Titov et al.(2011)]{titov11}  Titov, O., Lambert, S. B., \& Gontier, A.-M.  2011, A\&A, 529, A91
\bibitem[Titov \& Lambert(2013)]{titov13}  Titov, O. \&  Lambert, S.  2013, A\&A, 559, A95
\bibitem[Xu et al.(2012)]{xu12}  Xu, M. H., Wang, G. L., \& Zhao, M.  2012, A\&A, 544, A135
\end{thebibliography}
\end{document}